\documentstyle[prl,aps,multicol]{revtex}
\input epsf

\begin{document}
\title{The Meaning of the  Interaction-Free Measurements}

\author{ Lev Vaidman}

\maketitle
\vspace{.4cm}
 
\centerline{ School of Physics and Astronomy} 
\centerline{Raymond and Beverly Sackler Faculty of Exact Sciences}
\centerline{Tel-Aviv University, Tel-Aviv 69978, Israel}

\date{}

\vspace{.2cm}
\begin{abstract}
  Interaction-free measurements introduced by Elitzur and Vaidman
  [Found.  Phys.  {\bf 23}, 987 (1993)] allow finding infinitely
  fragile objects without destroying them.  Many experiments have been
  successfully performed showing that indeed, the original scheme and
  its modifications lead to reduction of the disturbance of the
  observed systems.  However, there is a controversy about the
  validity of the term ``interaction-free'' for these experiments.
  Broad variety of such experiments are reviewed and the meaning of
  the interaction-free measurements is clarified.
\end{abstract}
\vspace {.1cm}
\begin{multicols}{2}

\section{INTRODUCTION}
\label{intr}

The interaction-free measurements proposed by Elitzur and Vaidman
\cite{EV91,EV93} (EV IFM) led to numerous investigations and several
experiments have been performed
\cite{Kwi95,Voo97,HaSu97,Tse98,White98,Kwi981,Kwi99,MiMi99,Rud,ElDo01}.
One of the possible  applications of the interaction-free
measurements for quantum communication is that it opens up the way to
novel quantum non-demolition techniques \cite{en-ex2,Pot}. Other applications
are using the idea of interaction-free measurements for
``interaction-free'' computation \cite{Joz} and for improving
cryptographic schemes \cite{Guo,Cza}.  However, there have been
several objections to the name ``interaction-free''. Some authors in
trying to avoid it, made modifications such as ``interaction (energy
exchange) free measurements'' \cite{en-ex1,en-ex2}, ``indirect
measurements'' \cite{indi}, ``seemingly interaction-free
measurements'' \cite{seem-ifm}, ``interaction-free'' interrogation
\cite{Kwi99,Rud}, ``exposure-free imaging'' \cite{Ino},
``interaction-free interaction'' \cite{Hor01}, ``absorption-free
measurements'' \cite{MiMa}, etc.  Moreover, Simon and Platzman
\cite{SiPl} claimed that there is a ``fundamental limit on
`interaction-free' measurements''. In many works on the implementation and
the analysis of the EV IFM there is a considerable confusion about the
meaning of the term ``interaction-free''.  For example, a very recent
paper \cite{Pot} stated that ``energy exchange free'' is now well
established as a more precise way to characterize IFM in the case of
classical objects. On the other hand, Ryff and Ribeiro \cite{Ryff} used
the name ``interaction-free'' for a very different experiment. In this
paper I want to clarify in which sense the interaction-free
measurements are interaction free.  I will also make a comparison with
procedures termed ``interaction-free measurements'' in the past and
will analyze conceptual advantages and disadvantages of various modern
schemes for the IFM.

The plan of this paper is as follows: in Section II I will describe
the original proposal of Elitzur and Vaidman. Section III is devoted
to  a particular aspect of the IFM according to which the
measurement is performed without any particle being at the vicinity of
the measured object. The discussion relies on the analogy with the
``delayed choice experiment'' proposed by Wheeler \cite{Whee}. In
Section IV I make a comparative analysis of the ``interaction-free
measurements'' by Renninger and Dicke. In Section V I analyze
interaction-free measurements of quantum objects. Section VI devoted
to the controversy related to the momentum and energy transfer in the
process of the IFM. In Section VII I discuss modifications of the
original EV proposal, in particular, the application of the quantum
Zeno effect for obtaining a more efficient IFM. I end the paper with a
few concluding remarks in Section VIII.

\section{THE ELITZUR-VAIDMAN INTERACTION-FREE MEASUREMENTS}
\label{inter}

In the EV IFM paper the following question has been considered:
\begin{quotation}
    Suppose there is an object such that {\em any} interaction with it
  leads to an explosion. Can we locate the object without exploding
  it? 
\end{quotation}

The EV method is based on the Mach-Zehnder interferometer.  A photon
(from a source of single photons) reaches the first beam splitter
which has a transmission coefficient ${1\over2}$.  The transmitted and
reflected parts of the photon wave are then reflected by the mirrors
and finally reunite at another, similar beam splitter, see Fig.~1{\it a}.
Two detectors are positioned to detect the photon after it passes
through the second beam splitter.  The positions of the beam splitters
and the mirrors are arranged in such a way that (because of
destructive interference) the photon is never detected by one of the
detectors, say $D_2$, and is always detected by $D_1$.

This interferometer is placed in such a way that one of the routes of
the photon passes through the place where the object (an ultra-sensitive bomb) might be
present (Fig.~1{\it b}).  A single photon passes through the system.
There are three possible outcomes of this measurement: i)~explosion,~
ii)~detector $D_1$ clicks, iii)~detector $D_2$ clicks.  If 
detector $D_2$ clicks (the probability for that is ${1\over4}$), the
goal is achieved: we know that the object is inside the interferometer
and it did not explode.

The EV method solves the problem which was stated above. It allows
finding with certainty an infinitely
sensitive bomb without exploding it. The bomb might explode in the
process, but there is at least a probability of 25\% to find the bomb
without the explosion. ``Certainty'' means that when the process is successful ($D_2$
clicks), we know for sure that there is something
inside the interferometer. 

The formal scheme of the EV method is as follows. The first stage of
the process (the first beam
splitter) splits the wave packet of the test particle into
superposition of two wave-packets. Let us signify $|\Phi_{\rm
 int}\rangle$ is  the wave packet
which goes through the interaction region and   $|\Phi_{\rm
 free}\rangle$ is the wave packet which does not enter the interaction
region. In the basic EV procedure the first stage is
\begin{equation}
|\Phi\rangle \rightarrow {1\over {\sqrt 2}} (|\Phi_{\rm
 int}\rangle +|\Phi_{\rm free\rangle}).
\end{equation}

The next stage is the interaction between the object (the bomb) and 
the test particle. If the test particle enters the interaction region
when the bomb is present, it causes an explosion:
\begin{equation}
  |\Phi_{\rm  int}\rangle |{\rm bomb~in}\rangle 
 \rightarrow |{\rm explosion}\rangle .
\end{equation}

 If the test particle does not enter the interaction region or if the
 bomb is not  present, then nothing happens  at this stage:
\vskip .4cm

\begin{center} \leavevmode \epsfbox{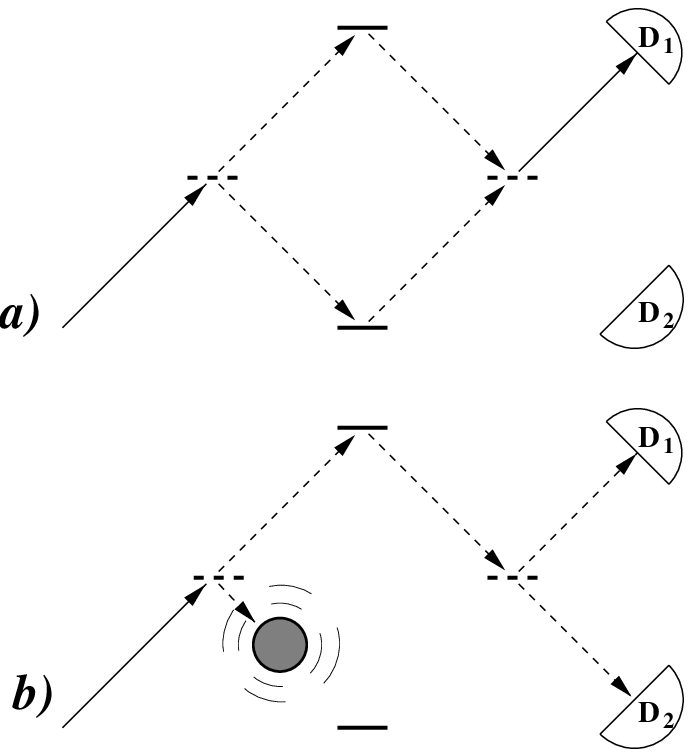} \end{center}
 
{\small  FIG. 1.~(a) When  the interferometer is
  properly tuned, all photons are detected by $D_1$ and none reach
  $D_2$. \hfill\break (b) If the bomb is present, detector $D_2$ has the
  probability 25\% to detect the photon  sent through the
  interferometer, and in this case we know that the bomb is inside the
  interferometer without exploding it.  }

\begin{eqnarray}
  \nonumber 
|\Phi_{\rm  free}\rangle |{\rm bomb~in}\rangle~\, 
& \rightarrow & |\Phi_{\rm  free}\rangle |{\rm bomb~in}\rangle ,\\
|\Phi_{\rm  int}\rangle |{\rm bomb~out}\rangle~
& \rightarrow & |\Phi_{\rm  int}\rangle |{\rm bomb~out}\rangle ,\\
 \nonumber
|\Phi_{\rm  free}\rangle |{\rm bomb~out}\rangle
& \rightarrow & |\Phi_{\rm  free}\rangle |{\rm bomb~out}\rangle .
\end{eqnarray}

The next stage is the observation of the interference between the two wave
packets of the test particle; it takes place at the second beam-splitter and
detectors. It is achieved by splitting the noninteracting wave packet
\begin{equation}
|\Phi_{\rm free\rangle} \rightarrow
 {1\over {\sqrt 2}} (|\Phi_1\rangle +|\Phi_2\rangle),
\end{equation}
and splitting the wave packet which passed through the interaction region
(if it did)
\begin{equation}
|\Phi_{\rm int\rangle} \rightarrow
 {1\over {\sqrt 2}} (|\Phi_1\rangle - |\Phi_2\rangle).
\end{equation}
The observation of the test particle is described by
\begin{eqnarray}
  \nonumber 
|\Phi_1\rangle  |D_1~{\rm  ready}\rangle |D_2~{\rm  ready}\rangle 
 \rightarrow
|\Phi_1\rangle  |D_1~{\rm  clicks}\rangle |D_2~{\rm  ready}\rangle , \\
|\Phi_2\rangle  |D_1~{\rm  ready}\rangle |D_2~{\rm  ready}\rangle 
 \rightarrow
|\Phi_2\rangle  |D_1~{\rm  ready}\rangle |D_2~{\rm  clicks}\rangle . 
\end{eqnarray}
The state  $|D_2~{\rm  clicks}\rangle$ corresponds to the success of the
experiment, when we know that the bomb is present in the interaction
region; we signify it by the state $|{\rm know~bomb~in}\rangle$. 

If the bomb is not present then the  the EV
measurement is described by
\begin{eqnarray}
\nonumber
|\Phi\rangle |{\rm bomb~out}\rangle |D_1~{\rm  ready}\rangle |D_2~{\rm
  ready}\rangle& 
\rightarrow ~~~~ \\
~~~~|\Phi_1\rangle |{\rm bomb~out}\rangle |D_1~{\rm  clicks}\rangle
|D_2~{\rm  ready}\rangle.& &
\end{eqnarray}
If the bomb is inside the interferometer,  then the  the EV
measurement is described by
\begin{eqnarray}
\nonumber
|\Phi\rangle |{\rm bomb~in}\rangle |D_1~{\rm  ready}\rangle |D_2~{\rm
  ready}\rangle  \rightarrow & \\
\nonumber
{1\over {\sqrt 2}}|{\rm explosion}\rangle + {1\over 2}|{\rm
  know~bomb~in}\rangle ~~~~&  \\
+ {1\over 2} |\Phi_1\rangle |{\rm bomb~in}\rangle |D_1~{\rm  clicks}\rangle
|D_2~{\rm  ready}\rangle . & 
\end{eqnarray}
 The experiment ends up in finding the bomb with the probability of
$25\%$, explosion with the probability of 
$50\%$, and no information but no explosion  with the probability
 of $25\%$. In the latter case we can repeat the procedure and in this
 way (by repeating again and again) we can find  one third of bombs
 without exploding them. It was found \cite{EV93} that changing the reflectivity
 of the beam splitters can improve the method such that the fraction
 of the  bombs remaining intact   almost reaches one half.

\section{Measurement without ``touching''}
\label{touch}

 The name ``interaction-free'' seems very appropriate for a procedure
 which allows finding objects without exploding them, in spite of the
 fact that these objects explode  due to {\it any} interaction.
  Simple logic tells us: given that any
 interaction leads to an explosion and given that there has been no
 explosion, it follows that there has been no interaction.  This
 argument which sounds unambiguous in the framework of classical
 physics requires careful definition of the meaning of ``any
 interaction'' in the domain of quantum mechanics.

The weakness of the definition: ``The IFM is a procedure which allows
finding an object exploding due to {\it any} interaction without
exploding it,'' is that quantum mechanics precludes existence of such
objects. Indeed, a good model for an ``explosion'' is an inelastic
scattering \cite{Gesz}. The Optical Theorem \cite{Land} tells us
that there cannot be an inelastic scattering without some elastic
scattering. The latter does not change the internal state of the
object, i.e., the object does not explode.  In order to avoid
non-existing concepts in the definition of the IFM, we should modify the
definition in the following way:
\begin{quotation}
The IFM is a procedure which allows
finding (at least
sometimes) bombs of any sensitivity without exploding them.
 \end{quotation}

The method presented in the EV IFM paper have
certain additional features which further justify the name
``interaction-free''. The method is applicable for finding the
location of objects which do not necessarily explode. Even for such an
object we can claim that, in some sense, finding  its location is
``interaction-free''.
The discussion about the justification of the term
``interaction-free'' for the EV procedure has started in the original
EV IFM paper \cite{EV93}:
  \begin{quotation}
   The argument which claims that this is an interaction-free
   measurement sounds very persuasive but is, in fact, an artifact of
   a certain interpretation of quantum mechanics (the interpretation
   that is usually adopted in discussions of Wheeler's delayed-choice
   experiment). The paradox of obtaining information without
   interaction appears due to the assumption that only one ``branch''
   of a quantum state exists. (p. 991)
 \end{quotation}
 One of the ``choices'' of Wheeler's delayed-choice experiment is an
 experiment with a Mach-Zehnder interferometer in which the second beam
 splitter is missing (see Fig.~2). In the run of the experiment with a
 single photon detected by $D_2$, it is usually accepted that the
 photon had a well defined trajectory: the upper arm of the
 interferometer. In contrast, according to the von Neumann approach,
 the photon was in a superposition inside the interferometer until the
 time when one part of the superposition reached the detector $D_2$ (or
 until the time the other part reached the detector $D_1$ if that
 event was earlier). At that moment the wave function of the photon
 collapses to the vicinity of $D_2$.  The justification of Wheeler's
 claim that the photon detected by $D_2$ never was in the lower arm of
 the interferometer is that, according to the quantum mechanical laws,
 we cannot see any physical trace from the photon in the lower arm of
 the interferometer. This is true if (as it happened to be in this
 experiment) the photon from the lower arm of the interferometer
 cannot reach the detector $D_2$. 

The fact that there cannot be
 a physical trace of the photon in the lower arm of the interferometer
 can be explained in the framework of the two-state vector formulation
 of quantum mechanics \cite{ABL,AV90}.  This formalism is particularly
 suitable for this case because we have pre- and post-selected
 situation: the photon was post-selected at $D_2$. While the wave
 function of the photon evolving forward in time does not vanish in the
 lower arm of the interferometer, the backward-evolving wave function
 does.  Vanishing one of the waves (forward or backward) at a
 particular location is enough to ensure that the photon cannot cause
 any change in the local variables of the lower arm of the
 interferometer.

\begin{center} \leavevmode \epsfbox{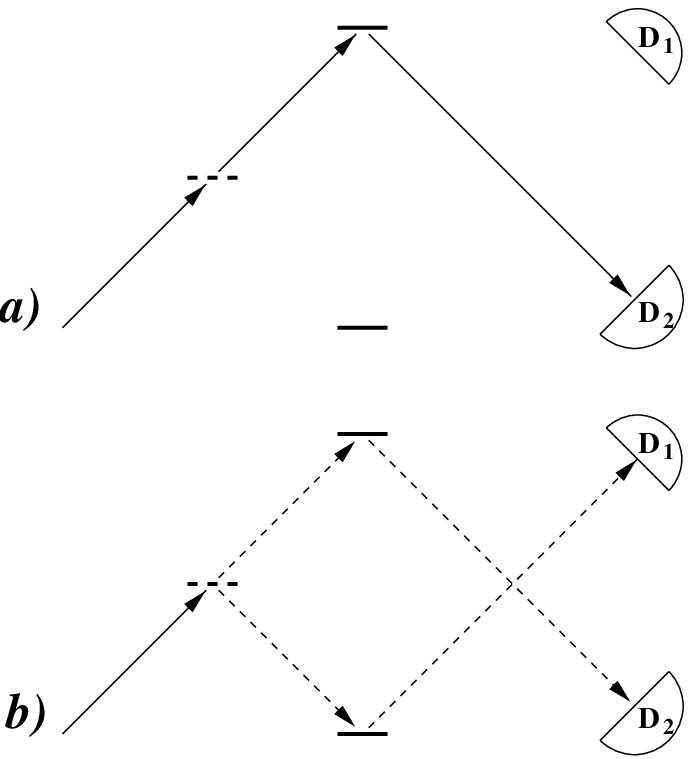} \end{center}

\noindent 
 {\small  FIG. 2.~ (a) The ``trajectory'' of the photon in the
   Wheeler experiment given that $D_2$ detected the photon, as it
   is usually described. The photon cannot leave any physical trace
   outside its ``trajectory''.  \hfill\break (b) The ``trajectory''
   of the quantum wave of the photon in the Wheeler experiment
   according to the von Neumann approach. The photon remains in a
   superposition until the collapse which takes place when one of the
   wave packets reaches a detector.  }

 \vskip .4cm
\break
 
 In our experiment (Fig.~1.) we have the same situation. If there is an object
 in the lower arm of the interferometer, the photon cannot go through
 this arm to the detector $D_1$. This is correct if the object is such
 that it explodes whenever the photon reaches its location, but moreover, this is also correct in the
 case in which the object is completely nontransparent and it blocks
 the photon in the lower arm eliminating any possibility of reaching
 $D_1$.  Even in this case we can claim that we locate the object
 ``without touching''. This claim is identical to the argument
 according to which the photon in Wheeler's experiment went solely
 through the upper arm.  In the framework of the two-state vector
 approach we can say that the forward-evolving quantum state is
 nonzero in the lower arm of the interferometer only up to the location of
 the object, while the backward-evolving wave function is nonzero only from
 the location of the object. Thus, at every point of the lower arm of
 the interferometer one of the quantum states vanishes. The
 two-state vector formalism  does not suggest that the photon is
 not present at the lower arm of the interferometer; it only helps to
 establish that the photon does not leave a trace there. The latter is
 the basis for the claim that, in some sense, the photon was not
 there.

\section{ The IFM of Renninger and Dicke}
\label{Re-Di}

In many papers describing experiments and modifications of the EV IFM
the first cited papers are one by Renninger \cite{Renn} and another by
Dicke \cite{Dick}. It is frequently claimed that Elitzur and Vaidman
``extended ideas of Renninger and Dicke'' or just ``amplified the
argument by inventing an efficient interferometric set'' \cite{Gesz}.
In fact, there is little in common between Renninger-Dicke IFM
and the EV IFM.  Dicke's paper is cited in the EV IFM paper, but the
citation is given only for the justification of the name:
``interaction-free measurements''. Renninger's and Dicke's papers do
not have the method, and, more importantly, they do not address the
question which the EV IFM paper have solved.

\begin{center} \leavevmode \epsfbox{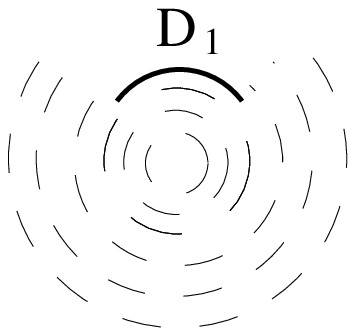} \end{center}

\vskip .1cm
{\small  FIG. 3.~ Renninger's experiment. The photon spherical wave is
  modified by  the scintillation detector $D_1$ in spite of the fact that
  it detects nothing.}
\vskip .3cm

\begin{center} \leavevmode \epsfbox{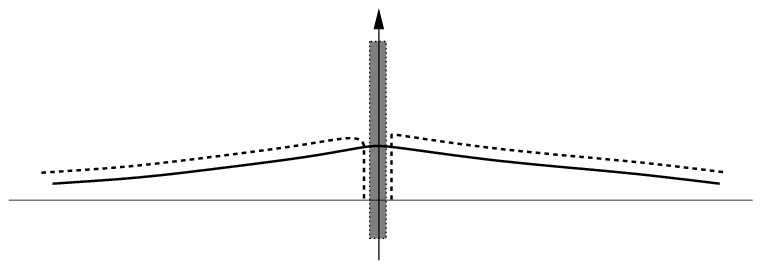} \end{center}

 {\small FIG. 4.~  Dicke's Experiment.~ The ground state of a
   particle in the potential well (solid line) is changed to a more energetic
   state (dashed line) due to short radiation pulse, while the quantum state of
   the photons in the pulse remains unchanged.}
\vskip .3cm

Renninger discussed a {\it negative result experiment}: a situation in
which the detector does not detect anything.  In spite of the fact
that nothing happened to the detector, there is a change in the
measured system. He considered a spherical wave of a photon after it
extended beyond the radius at which a scintillation detector was located
in  part of the solid angle, see Fig.~3. The state of the detector
remained unchanged but, nevertheless, the wave-function of the photon
is modified. The name ``interaction-free'' for Renninger's setup might
be justified because there is not {\it any}, not even an infinitesimally
small, change in the state of the detector in the described process.
This is contrary to the classical physics in which interaction in
a measurement process can be made arbitrary small, but it cannot be
exactly zero.

Dicke considered the paradox  of the apparent non-conservation of
 energy in a Renninger-type experiment. He considered an atom in a
 ground state inside a potential well. Part of the well was
 illuminated by a beam of photons. A negative result experiment was
 considered in which no scattered photons were observed, see Fig.~4.
 The atom changed its state from the ground state to some 
 superposition of energy eigenstates (with a larger expectation value of
 energy) in which the atom does not occupy the part of the well
 illuminated by the photons. The  photons, however, 
 apparently have not changed their state at all. Then, Dicke asked: ``What is
 the source of the additional energy of the atom?!''

 Careful analysis \cite{Dick2,Gold} (in part, made by Dicke himself)
 shows that there is no real paradox with the conservation of energy,
 although there are many interesting aspects in the process of an
 ideal measurement \cite{Pearl}. One of the key arguments is that the
 photon pulse has to be well localized in time and, therefore, it must
 have a large uncertainty in energy.

The word ``measurement'' in quantum theory have many very different
meanings \cite{Bell}.  The purpose of the Renninger and Dicke measurements
is {\it preparation} of a quantum state. In contrast, the purpose of
the EV interaction-free measurement is to obtain {\it information}
about the object. In Renninger and Dicke measurements the {\it
  measuring device} is undisturbed (these are negative result
experiments) while in the EV measurement the {\it observed object} is,
in some sense, undisturbed.  In fact, in general EV IFM the quantum
state of the observed object {\it is} disturbed: the wave function
becomes localized at the vicinity of the lower arm of the
interferometer (see Sec. 3 of the EV paper). The reasons for using the
term ``interaction-free measurements'' are that the object does not
explode (if it is a bomb), it does not absorb any photon (if it is an
opaque object) and that we can claim that, in some sense,  the photon
does not reach the vicinity of the object.

A variation of Dicke's measurement which can serve as a measurement
of the location of an object was considered in the EV IFM paper for
justifying the name ``interaction-free measurements'' of the EV procedure.
An object in a superposition of being in two far away places was
considered. A beam of light passed through one of the locations and no
scattered photons were observed. This yields the information that the object
is located in the other place.  The described experiment is interaction-free
because the object (if it is a bomb) would not explode: the object is
found in the place where there were no photons.

 In such an experiment,
however, it is more difficult to claim that the photon was not at the
vicinity of the object: the photon was not at the vicinity of the {\it
  future} location of the object. But the main weakness of this
experiment relative to the EV scheme is that we get information about
the location of the object only if we have {\it prior information} about
the state of the object. If it is known in advance that the object can
be found in one  of two boxes and it was not found in one,
then obviously, we  know that it is in the second box.  The whole
strength of the EV method is that we get information that an object is
inside the box {\it without any prior information!}  The latter,
contrary to the former task cannot be done without help of a quantum
theory.

In order to see the difference more vividly 
 let us consider an application of the EV method to
 Dicke's experimental setup.  Instead of the light pulse we send a
 ``half photon'': We arrange the EV device such that  one arm of the
 Mach-Zehnder interferometer passes through the location of the
 particle, see Fig.~5.  Then, if detector $D_2$
 clicks, the particle is localized in the interaction region.
 
 In both cases (the Renninger-Dicke IFM and this EV IFM) there is a
 change in the quantum state of the particle without, in some sense,
 interaction with  the photon.  However, the situations are quite
 different. In the original Dicke's experiment we can claim that the
 dashed line of Fig.~4.  is the state of the particle after the
 experiment only if we have prior information about the state of the
 particle before the experiment (solid line of Fig.~4.) In contrast,
 in the EV modification of the experiment, we can claim that a
 particle is localized in the vicinity of the interaction region (dashed
 line of Fig.~5.) even if we had no prior information about the
 state of the particle.

It seems that Dicke named his experiment ``interaction-free'' mainly
because  the photons did not scatter: this is a ``negative result
experiment''. In the EV experiment the photon clearly changes its
state and it is essential that it was detected: this is not a ``negative
result experiment'' in this sense.

Paul \cite{Paul} noted that there is an earlier paper by Renninger \cite{Ren53} in which   an experimental setup 
almost identical to that of the EV IFM was considered: a Mach-Zehnder
interferometer tuned to have a dark output towards one of the
detectors. However, Renninger never regarded his experiment as a
measurement on an object which was inside the interferometer:
Renninger's argument, as in the experiment described in Fig. 3, was
about ``interaction-free'' changing the state of the photon. Renninger
has not asked the key question of the EV IFM: How to get information
in an interaction-free manner?

 I can see something in common between the Renninger-Dicke IFM and the EV
IFM  in the framework of the many-worlds interpretation.  In both
cases there is an  ``interaction'': radiation of the scintillator in
the Renninger experiment or explosion of the bomb in the EV experiment,
but these interactions take place in the ``other'' branch, not in the
branch we end up discussing the experiment. In an attempt to avoid
adopting the many-worlds interpretation such interactions were considered as
{\it counterfactual} \cite{Pen,Mi-Jo}.

\begin{center} \leavevmode \epsfbox{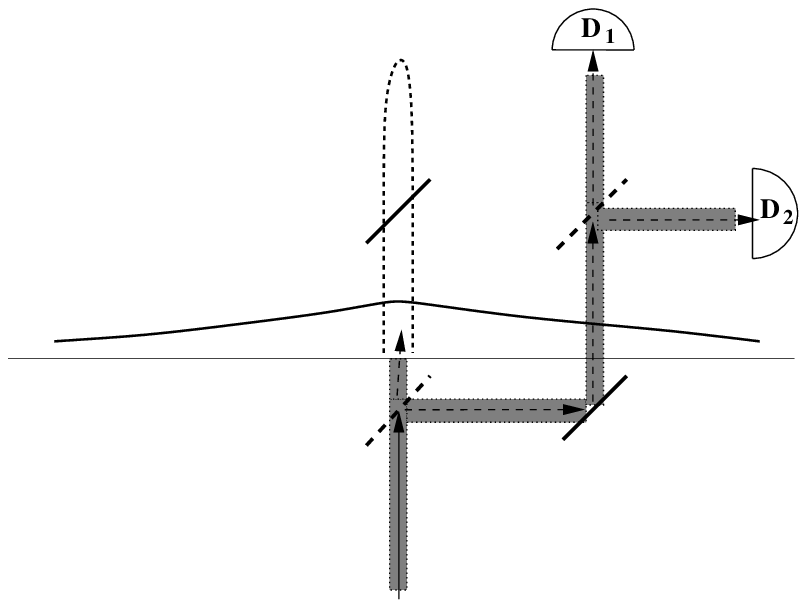} \end{center}

\noindent 
{\small FIG. 5.~  The EV modification of Dicke's
  Experiment.~ The ground state of a particle in the potential well (solid
  line) is changed to a well  localized state (dashed line) when the
  photon is detected by the detector $D_2$.}

\section{ Interaction-free localization of a quantum object}
\label{IFML}
 
We name  the experiment described in Fig.~5. ``interaction-free''
measurement (cf. ``interaction-free collapse'' of the  EV IFM paper)
in spite of  the fact that both the particle and the photon change
their states. The main motivation for the name is that the
interaction between the particle and the photon is such that there is
an ``explosion'' if they ``touch'' each other, but the experiment (when
$D_2$ clicks) ends up without explosion.

 The second aspect of the EV IFM, when applied to quantum objects, encounters
a  subtle difficulty. After performing the procedure of the IFM and
obtaining the photon click at $D_2$, we cannot claim that the photon
was not present at the region of interaction; moreover, it might be the
case that, in some sense, the photon was there  {\it  with certainty}.

First, let us repeat the argument which led us to think that the
photon was not there. Consider again the experiment described on Fig.
1., but now the ``bomb'' is replaced by a quantum object in a
superposition of being in the ``interaction region'' and somewhere else
outside the interferometer. If $D_2$ clicks, we can argue that the
object  had to be on the way of the photon in the lower arm of the
interferometer, otherwise, it seems that we cannot explain the arrival
of the photon to the ``dark'' detector $D_2$. If the object was on
the way of the photon, we can argue that the photon was not there,
otherwise we had to see the explosion. Therefore, the photon went
through the upper arm of the interferometer and it was not present in
the interaction region.

The persuasive argument of the previous paragraph is incorrect! Not
just the semantic point discussed above, i.e., that according to the
standard approach the quantum wave of the photon in the lower arm of
the interferometer was not zero until it reached the interaction
region. It is wrong to say that the photon was not in the lower arm
even in the part {\it beyond} the interaction region.  In the
experiment in which $D_2$ clicks, the photon {\it can} be found in any
point of the lower arm of the interferometer!

This claim can be seen most clearly by considering ``nested
interaction-free measurements'' \cite{Hardy}.  The object is in a
superposition of two wave packets inside its own Mach-Zehnder
interferometer (see Fig.~6.)  If $D_2$ (for the photon) clicks, the
object is localized inside the interaction region $W$.  However,
the object itself is the test particle of another IFM (we can consider
a gedanken situation in which the object which explodes when the
photon reaches its location can, nevertheless, be manipulated by other
means). If this other IFM is successful (i.e.  ``$D_2$'' for the
object clicks) then the other observer can claim that she localized
the photon of the first experiment at $W$, i.e. that the photon passed
through the lower arm of the interferometer on its way to $D_2$.

\begin{center} \leavevmode \epsfbox{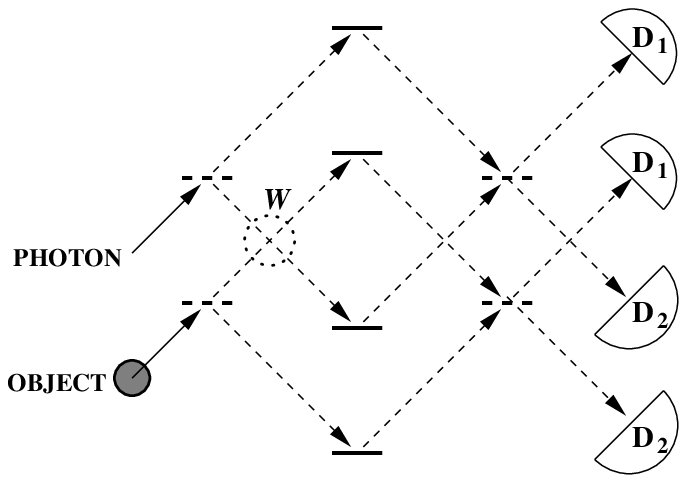} \end{center}
 
{\small  FIG. 6.~  Hardy's Paradox. Two interferometers
  are tuned in such a way that, if they operate separately,  there
  is a complete destructive interference towards detectors $D_2$. The
  lower arm of the photon interferometer intersects the upper arm of
  the object interferometer in $W$ such that the object and the
  photon cannot cross each other.  When the photon and the object
  are sent together (they reach $W$ at the same time) then there is a
  nonzero probability for clicks of both detectors $D_2$. In this case
  one can infer that the object was localized at $W$ and also that
  the photon was localized at $W$.  However, the photon and the
  object were not present in $W$ together. This apparently
  paradoxical situation does not lead to a real contradiction because
  all these claims are valid only if tested separately.}

\vskip .3cm

Paradoxically, all these claims are true (in the operational sense):
if we look for the photon in $W$, we find it with certainty; if we
look, instead, for the object in $W$, we find it with certainty too.
Both claims are true separately, but not together: if we look for the
pair, the photon and the object together, in $W$, we fail with
certainty.  Such peculiarities take place because we consider a pre-
and post-selected situation (the post-selection is that in both
experiments detectors $D_2$ click) \cite{Har-Vai}. An interesting
insight about this peculiar situation can be learned through the
analysis of the {\it weak measurements} performed on the object and
the photon inside their interferometers \cite{Ha-wea}.

In spite of this peculiar feature, the experiment is still
interaction-free in the following sense.  If somebody would test the
success of our experiment for localization of the object, i.e. would
measure the location of the object shortly after the ``meeting time''
between the object and the photon, then we know with certainty that she
would find the object in $W$ and, therefore, the photon cannot be
there. Discussing the issue of the presence of the object with her, we
can correctly claim that in our experiment the photon was not in the
vicinity of the object. Indeed, given the assumption that she found
the object, we know that  she has not seen the
photon in the lower arm of the interferometer, even if she looked for it there.
  However, if, instead of measuring the position of the
object after the meeting time, she finds the object in a particular
superposition (the superposition which with certainty reaches $D_2$),
she can claim with certainty that the photon was in $W$. (Compare this
with {\it deterministic quantum interference experiments} \cite{APP}).

\section{Momentum and energy  transfer in the IFM}
\label{trans}

Probably, the largest misconception about the IFM is defining them as
momentum and energy exchange-free measurements \cite{en-ex1,en-ex2,SiPl}. The EV IFM can
localize a bomb in an arbitrary small region without exploding it even
if the quantum state of the bomb was spread out initially.
Localization of an object without uncertain change in its momentum
leads to immediate contradiction with the Heisenberg uncertainty
principle. Identifying the interaction-free measurements as
momentum-exchange free measurements, Simon and Platzman \cite{SiPl} derived
``fundamental limits'' on the IFM. They argued that the IFM can be
performed only on infinitely sensitive bomb and that a bomb which is
infinitely sensitive to any momentum transfer could not be placed in
the vicinity of the IFM device from the beginning. These arguments
fail because the EV IFM are not defined as momentum-exchange free
measurements. (Probably, the misconception came because of frequent
mentioning of Dicke's paper \cite{Dick} which concentrated on the issue of the
energy exchange in his IFM.)

The arguments, similar to those of Simon and Platzman might be relevant for
performing a modification of the EV IFM proposed by Penrose
\cite{Pen}. 
 He proposed a method  for testing some property of an object
without interaction. The object is again a bomb which explodes when
anything, even a single photon, ``touches'' its trigger device. Some of the
bombs are ``duds'': their trigger
device is locked to a body of the bomb and no explosion and no relative motion
of the trigger device would happen when it is ``touched''. 
Again, the paradox is that any touching of a trigger of a good bomb leads to an
explosion, but, nevertheless,  good bombs can be found (at least
sometimes) without the explosion.

In the Penrose version of IFM, the bomb plays the role of one mirror of
the interferometer, see Fig.~7. It has to be placed in the correct
position. We are allowed to do so by holding the body of the bomb. However, the
uncertainty principle puts limits on placing the bomb in its place
before the experiment \cite{Pen-Vai}. Only if the position of the bomb
(in fact, what matters is the position of the dud) is known exactly, the
limitations are not present.  In contrast, in the EV IFM the bomb need
not be localized prior to the measurement: the IFM localizes it by
itself.

\begin{center} \leavevmode \epsfbox{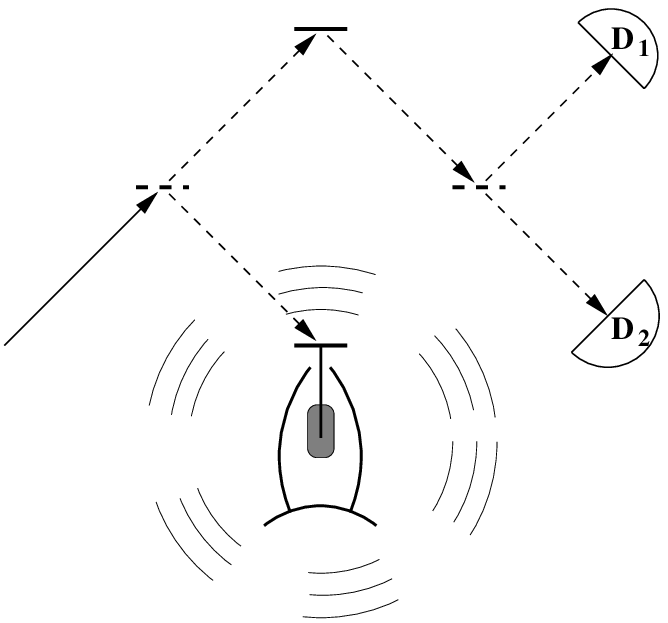} \end{center}

{\small  FIG. 7.~  The Penrose bomb-testing device.~ The mirror of the good bomb
  cannot reflect the photon, since the incoming photon causes an
  explosion. Therefore, $D_2$ sometimes clicks. The mirror of a dud
  is connected to the massive body, and therefore the interferometer
  ``works'', i.e. $D_2$ never clicks when the mirror is a dud.  }

\vskip .3cm

The zero change in the momentum of the object, location of which is
found in the IFM, is not a necessary condition for the measurement to
be IFM, but there are IFM in which  there is no  change of the  momentum of the object.
  Indeed, if  the object has been  localized before the IFM procedure,
then its state and, therefore, its momentum distribution do not change
during the process.  

 The relevant issue seems to be the change in the momentum of the observed
object, but it is interesting to consider also the change in the momentum
of the measuring device, thus analyzing  the question of the {\it
  exchange} of the momentum. If the object is localized from the
beginning then its state does not change, but the state of the photon
does change: from the superposition of being in two arms of the
interferometer it collapses into a localized wave packet in one arm of
the interferometer. It can be arranged that the two separate wave
packets of the photon have the same distribution of momentum. Then,
the collapse to one wave packet will not change expectation value of
any power of momentum of the photon.


Aharonov \cite{Aha-pri} has pointed out that although in this process
there is no exchange of momentum in the above sense, still there is an
exchange of certain physical variable. In the EV procedure there is an
exchange of {\em modular momentum}. The collapse of the quantum wave
of the photon from the superposition of the two wave packets separated
by a distance $a$ to a single wave packet  is accompanied by the change in the
modular momentum  $p_{phot} {\rm mod}{\hbar \over a}$.
The modular momentum of the object localized at the lower arm of the interferometer from
the beginning, $p_{obj} {\rm mod}{\hbar \over a}$,
  does not change (there is no {\em any} change in the
quantum state of the object). One can, nevertheless,  consider an 
exchange of modular momentum in this process: since $p_{obj} {\rm
  mod}{\hbar \over a}$ is completely uncertain, there is no 
contradiction with the conservation law for the total modular
momentum.

Note that the situation in which the expectation values of
any power of momentum remains unchanged, while expectation values of
powers of modular momentum change, is also a  feature of Aharonov-Bohm type effects in which  the quantum state changes even though no local
forces are acting.

The method of the EV IFM can be applied for performing various
non-demolition measurements \cite{en-ex2}. Indeed, even if the
measurement interaction can destroy the object, the method allows
measurement without disturbing the object.  However, not {\it any}
non-demolition measurement is an IFM in the sense I discussed it here.
In some nondemolition experiments the test particle of the measuring
device explicitly passes through the location of the measured object.
In other experiments the state of the object changes, but these
changes are compensated at the end of the process \cite{Aha-comp}. I
suggest that such measurements should not be considered as
interaction-free.

\section{ Modifications of the EV IFM}
\label{mod-ifm}

The optimal scheme presented in the IFM paper allows detection of
almost 50\% of the bombs without explosion (the rest explode in the
process).  Kwiat {\it et al.} \cite{Kwi95} applied quantum Zeno effect
for constructing the IFM scheme which, in principle, can be made
arbitrary close to the 100\% efficiency. The experiment with
theoretical efficiency higher than 50\% has been performed
\cite{White98}.

The almost 100\% efficient scheme of Kwiat {\it et al.} \cite{Kwi95}
can be explained as follows. The experimental setup consists of two
identical optical cavities coupled through a highly reflective mirror,
see Fig.~8. A single photon initially placed in the left cavity. If
the right cavity is empty, then, after a particular number $N$ of
reflections, the photon with certainty will be in the right cavity. If,
however, there is a bomb in the right cavity, the photon, with the
probability close to 1 for large $N$, will be found in the left cavity.
Testing at the appropriate time for the photon in the left cavity,
will tell us if there is a bomb in the right cavity.

This method keeps all conceptual features of the EV IFM. If the photon
is found in the left cavity, we are certain that there is an object in
the right cavity. If the object is an ultra-sensitive bomb or if it is
completely non-transparent object which does not reflect light
backwards (e.g., it is a mirror rotated by $45$ degrees relative to
the optical axes of the cavity as in the Kwiat {\it et al.} experiment)
then, when we detect the photon in the left cavity we can claim that
it never ``touched'' the object in the same sense as it is true in the
original EV method.

\begin{center} \leavevmode \epsfbox{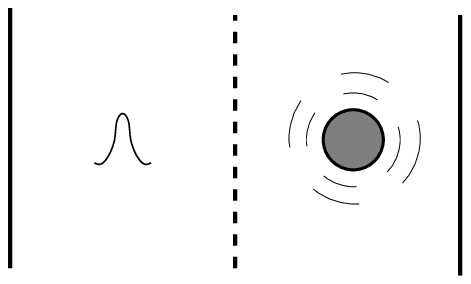} \end{center}

{\small  FIG. 8.~ The almost 100\% efficient scheme of the IFM.
If there is a ``bomb'' or a nontransparent object in the right cavity, 
then the photon stays in the left cavity, with  a
 probability to go to the right cavity which can be made arbitrary
 small by increasing the reflectivity of the mirror between the
 cavities. If, however, the right cavity is empty, then after some time the photon will
 move there with certainty.  }

\vskip .4cm

Another modification of the EV IFM which leads to the efficiency of
almost 100\% has been proposed by Paul and Pavi\v ci\'c \cite{PaPa}
and implemented in a laboratory by Tsegaye {\it et al.}  \cite{Tse98}.
The basic ingredient of this method is an optical resonance cavity
which is almost transparent when empty, and is an almost perfect
mirror when there is an object inside.  The advantage of the proposal
of Paul and Pavi\v ci\'c is that it has just one cavity, and is easier
to perform.  In fact, this method has been recently applied for
``exposure-free imaging'' of a two-dimensional object \cite{Ino}.
However, one cavity method has a conceptual drawback. In this
experiment there is always a nonzero probability to reflect the photon
even if the cavity is empty.  Thus, detecting reflected photon cannot
ensure presence of the object with 100\% certainty. Essentially, this drawback has
only an academic significance. In any real experiment there will be
uncertainty anyway, and the uncertainty which I mentioned can be
always reduced below the level of the experimental noise.

Other modifications of the IFM are related to interaction-free
``imaging''\cite{Kwi99} and interaction-free measurements of
semi-transparent objects \cite{Jang,MiMa}.  These experiments hardly
pass the strict definition of the IFM in the sense that the photons do
not pass in the vicinity of the object. However, they all achieve a
very important practical goal, since we ``see'' the object reducing  very
significantly the irradiation of the object: this can allow measurements
on fragile objects. Indeed, in spite of the fact that for distinguishing
small differences in the transparency of an object the method is not
very effective \cite{Krenn,MMP}, it still can be useful for reduced
irradiation pattern recognition \cite{KW}.

Reasoning in the framework of the many-worlds interpretation (MWI)
\cite{Va-par} leads to the statement that while we can find an object
in the interaction-free manner, we cannot find out that a certain
place is empty in the interaction-free way. Here, I mean
``interaction-free'' in the sense that no photons (or other particles)
pass through the place in question.  Getting information about some
location in space without any particle being there is paradoxical
because physical laws include only local interactions. In the case of
finding the bomb, the MWI solves the paradox. Indeed, the
laws apply to the whole physical Universe which includes all the
worlds and, therefore, the reasoning must be true only when we consider all the
worlds. Since there are worlds with the explosion we cannot say on the
level of the physical Universe that no photons were at the location of
the bomb. In contrast, when there is no bomb, there are no other
worlds. The paradox in our world becomes the paradox for the whole
Universe which is a real paradox. Thus, it is impossible to find a
procedure which tests the presence of an object in a particular place
such that no particles visit the place both in the case the object is
there and in the case the object is not there. Quantitative analysis
of the limitations due to this effect were recently performed by Reif
who called the task ``interaction-free sensing'' \cite{Reif}. This
effect also leads to limitations on the efficiency of
``interaction-free computation'' when all possible outcomes are
considered \cite{Mi-Jo}.

\section{ Conclusions }
\label{conc}

I have reviewed various analyses, proposals, and experiments of IFM
and measurements based on the EV IFM method. The common feature of 
these proposals is that we obtain information about an object while
significantly reducing its irradiation. 

The meaning of the EV IFM is that if an object changes its internal
state (not the quantum state of its center of mass) due to the
radiation, then the method allows detection of the location of the
object without {\it any} change in its internal state. There is no any
fundamental limit on such IFM. The IFM allow measurements of position
of infinitely fragile objects. In some sense it locates objects
without ``touching'', i.e. without particles of any kind passing
through its vicinity. I have clarified the limited validity of this
feature for IFM performed on quantum objects.

 Numerous papers on the IFM interpreted the concept of
 ``interaction-free'' in many different ways. I hope that in this work I
 clarified the differences and stated unambiguously the meaning of the
 original proposal.

\vspace{.3cm}
 \centerline{\bf  ACKNOWLEDGMENTS}
 
 It is a pleasure to thank Yakir Aharonov, Berge Englert, 
 and Philip Pearle for helpful discussions.
 This research was supported in part by grant 471/98 of the Basic
 Research Foundation (administered by the Israel Academy of Sciences
 and Humanities) and the EPSRC grant  GR/N33058.

\end{multicols}

\end{document}